\documentclass[sigconf]{acmart}

\AtBeginDocument{%
  \providecommand\BibTeX{{%
    \normalfont B\kern-0.5em{\scshape i\kern-0.25em b}\kern-0.8em\TeX}}}

\setcopyright{iw3c2w3}
\copyrightyear{2020}
\acmYear{2020}

\begin{document}

\title{VLEngagement: A Dataset of Scientific Video Lectures for\\Evaluating Population-based Engagement}


\author{Sahan Bulathwela, Mar\'ia P\'erez-Ortiz, Emine Yilmaz and John Shawe-Taylor}
\email{m.bulathwela@ucl.ac.uk}
\affiliation{%
 \institution{Centre for Artificial Intelligence, University College London}
 \streetaddress{Gower Street}
 \city{London, UK}
 \postcode{WC1E 6BT}
}


 

\renewcommand{\shortauthors}{Bulathwela, et al.}

\begin{abstract}
 With the emergence of e-learning and personalised education, the production and distribution of digital educational resources have boomed. Video lectures have now become one of the primary modalities to impart knowledge to masses in the current digital age. The rapid creation of video lecture content challenges the currently established human-centred moderation and quality assurance pipeline, demanding for more efficient, scalable and automatic solutions for managing learning resources. Although a few datasets related to engagement with educational videos exist, there is still an important need for data and research aimed at understanding learner engagement with scientific video lectures. This paper introduces VLEngagement, a novel dataset that consists of content-based and video-specific features extracted from publicly available scientific video lectures and several metrics related to user engagement. We introduce several novel tasks related to predicting and understanding context-agnostic engagement in video lectures, providing preliminary baselines. This is the largest and most diverse publicly available dataset to our knowledge that deals with such tasks. The extraction of Wikipedia topic-based features also allows associating more sophisticated Wikipedia based features to the dataset to improve the performance in these tasks. The dataset, helper tools and example code snippets are available publicly at \url{https://github.com/sahanbull/context-agnostic-engagement}
.
\end{abstract}



\maketitle

\section{Introduction}

Formal evaluations have shown that intelligent tutoring systems produce similar learning gains as one-on-one human tutoring, which has the potential to increase student performance to around the 98 percentile in a standard classroom \cite{woolf2010,Corbett2001, Bloom84}. Additionally, intelligent tutors could effectively reduce by one-third to one-half the time required for learning \cite{woolf2010}, increase effectiveness by 30\% as compared to traditional instruction \cite{woolf2010, fletcher1996does, fletcher1988intelligent}, reduce the need for training support personnel by about 70\% and operating costs by about 92\% and facilitate education in developing countries \cite{nye2015intelligent,vinuesa2020role}. Thus, the idea of building intelligent tutoring systems that provide online personalised education has gained a lot of traction in the recent years and will continue to do so. 

With more learning resources being created every day, automatic, scalable tools for quality assurance become essential \cite{camilleri}. Large educational resource repositories need scalable tools to understand/ estimate the engagement potential of newly added materials before exposing it to the learner audience \cite{clementsQ}. Thus, estimating context-agnostic (also named here population-based) engagement of materials and releasing related datasets becomes a critical part of quality assurance, recommendation and information retrieval. 

This work presents VLEngagement, a novel dataset that covers over 4000 peer-reviewed scientific video lectures constructed from a popular OER repository, VideoLectures.NET. The dataset has been proven very useful in some of our previous work, spanning both applications related to personalised and population based educational recommender systems. We believe the dataset has incredible potential for building intelligent educational and scientific recommender systems, specially given that similar datasets are usually proprietary and not publicly available. The dataset provides an extensive set of textual and video-specific features extracted from the lecture transcripts, together with Wikipedia topics covered in the lecture (via entity linking) and user engagement labels for each lecture. The dataset covers a wide set of tasks that are of crucial importance to building intelligent tutors, scalable quality assurance and understanding the features involved in population-based engagement.
While video retrieval and ranking are actively researched areas, this dataset allows adapting algorithms specifically for scientific video content which is novel and critical to building information retrieval systems in education and science.

The dataset is particularly suited to solve the cold-start problem found in educational recommender systems, both when i) new users join the system and we may not have enough information about their context so we may simply recommend population-based engaging lectures for a specific query topic and ii) new educational content is released, for which we may not have user engagement data yet and thus an engagement predictive model would be necessary. To the best of our knowledge, this is the first dataset to tackle such a task in education/scientific recommendations. The aim of the dataset is not to replace personalised recommenders by building population-based models, but rather to enable mixing personalised approaches with a meaningful population baseline/prior to solve the common cold-start problem and, given the context of the learner, rank the suitable material by their engagement potential \cite{trueeducation}.

\section{Related Work}

The learning analytics and educational data mining communities have developed novel algorithms to trace knowledge in learners \cite{vie2019knowledge,Piech2015} and provide personalised recommendations for educational material \cite{truelearn,pardos_serendipity}. All these approaches are focused on capturing and exploiting the \emph{context} of the learner. This is, among others, the knowledge state, interests, preferences and learning goals, all of which are crucial variables to develop an effective personalised tutoring system. However, in a landscape where new educational resources are created and circulated at a rapid scale (e.g. Open Educational Resources (OER) \cite{unesco1} and Massively Open Online Courses \cite{ramesh2014learning}), there is a big gap of knowledge in our understanding of the features involved in \emph{context-agnostic} (i.e. population-based) engagement of educational resources and the relationship between different measures of engagement, popularity and subjective assessment in general. Our work thus focuses on connecting content analytics to population-based learning analytics.

Learner engagement is a necessary prerequisite for acquiring knowledge from educational resources. \citeauthor{carini2006student} has shown that student engagement positively correlates with desired learning outcomes such as critical thinking and better grades in a conventional classroom setting. Various studies have demonstrated that learner engagement plays a key role in successful achievement of expected learning outcomes in a online learning setting \cite{slater2016semantic,ramesh2014learning}. \emph{Engagement} is a loaded concept that can have different definitions to different communities. For example, engagement is measured using different metrics depending on the modality of the educational resource. 

Most work related to modelling educational engagement attempts to model engagement as a function of the context of the learner  \cite{bonafini2017much,stud_eng_ou,beal2006classifying}. Our work, on the other hand, proposes to model context-agnostic engagement through several content-based features of the educational resource.
Context-agnostic engagement has been previously studied for video lectures, albeit from a more qualitative perspective, with general recommendations such as keeping videos short \cite{Guo_vid_prod} and using conversational language for lecture delivery \cite{brame2016effective}.
These recommendations help authors to create better educational videos, but none of these works address the need for predicting automatically highly engaging educational resources, which is crucial for retrieving and recommending educational material at scale.

\subsection{Related Datasets}

The interest in identifying engaging information goes beyond the educational domain and is investigated in numerous other fields. These works show that numerous feature verticals associated to content, such as \emph{understandability, freshness, topic coverage, presentation} and \emph{authority} exist \cite{quality_features}.  
Engagement (specifically watch time) has been used as the main measure used
 for YouTube recommendations \cite{Covington2016} and to predict engagement with general-purpose videos \cite{beyondviews}. This is usually the case for most media recommenders and several datasets are available for such task. 
 
 Looking beyond videos, Wikipedia uses a review system to evaluate the quality of its articles and several attempts have been made to build machine learning predictive models using
 features such as text style, readability, structure, network, recency and review information \cite{Dalip_wiki_svr,wiki_wang}.
 This Wikipedia article quality dataset \cite{Dalip_wiki_svr} is publicly available although user engagement data is not included. Only explicit quality labels are provided. Similar datasets are available for automated essay scoring \cite{taghipour2016neural}.
 
 There are only a handful of publicly available datasets that are related to predicting engagement in videos. 
Large-scale datasets focused on predicting engagement with general purpose videos (such as the one in \cite{beyondviews} that analyses engagement in YouTube) are common but these lack focus on educational material. Some of the features used by these works share some similarity to the ones used in this paper (such as video duration, category, language and topic features). However, a large part of the features are focused on the reputation of the YouTube channel. No textual features relating to understandability and presentation are used, since this may be of less importance for general purpose videos than for education.

The most relevant dataset in the literature to understanding engagement in video lectures, i.e. the work based on  approximately 800 videos from a Massively Open Online Course (MOOC) \cite{Guo_vid_prod}, is not publicly available. In this case, the data was manually processed and authors provided a qualitative analysis of engagement, with some features being relatively subjective and difficult to automate. 
A similar work \cite{edx_quality} takes 22 EdX videos, extracts cross-modal features and manually annotates their quality. This dataset is also not publicly available and does not focus on learner engagement or subjective assessment metrics. 

Although online learning platforms such as {EdX} \cite{Guo_vid_prod,edx_quality}, {Khan Academy} \cite{khan_bigdata} and other platforms harvest valuable learner behavioural data that is created in an "in-the-wild" setting, the datasets are often not publicly released due to the proprietary nature of the content and the user data. This work addresses this significant gap of data by constructing and releasing a dataset with over 4,000 scientific video lectures (OERs) associated with explicit star ratings and implicit engagement signals from hundreds of thousands of informal learners consuming video lectures in an in-the-wild setting.

Finally, a different line of work focuses on studying and identifying engagement from the learners perspective, through the recording of user learning sessions and brainwaves \cite{multa_dataset} and the use of computer vision and affect recognition \cite{kaur2018prediction} to propose \emph{automatic, semi-automatic} and \emph{manual} techniques. Although multiple datasets exist to address this task, these datasets are usually collected in lab setting using a limited number of participants \cite{dewan2019engagement}. 
However, the focus of these datasets is to detect learner engagement using a set of multi-modal data related to the learner (brain waves, visual information, learning logs, etc.) rather than the features of the content itself. 
There are also a handful of recent public datasets/competitions relating to how students interact with learning problems (e.g. assistments\footnote{\url{https://sites.google.com/site/assistmentsdata/home/assistment-2009-2010-data}} or multiple choice questions\footnote{\url{https://www.microsoft.com/en-us/research/event/diagnostic-questions-neurips2020/}}), but these datasets do not focus on engagement. 
 






\section{VLEngagement Dataset}

The VLEngagement dataset is constructed using the aggregated video lectures consumption data coming from a popular scientific OER repository, VideoLectures.Net\footnote{{\url{www.videolectures.net}}}. These videos are recorded when researchers are presenting their work at peer-reviewed conferences. Lectures are thus
reviewed and material is controlled for correctness of knowledge.  
It is noteworthy that the dataset consists of \emph{scientific video lectures} that explain novel scientific work geared more towards postgraduate, PhD level learners and the scientific research community. Therefore, the learner audience of the video lectures in this dataset may significantly differ from one of a conventional MOOC platform.

The dataset provides a set of statistics aimed at studying population based engagement in video lectures, together with other conventional metrics in subjective assessment such as average star ratings and number of views. We believe the dataset will serve the community applying AI in Education to further understand what are the features of educational material that makes it engaging for learners.

\subsection{Feature Extraction}
The dataset provides three types of features as outlined in Table \ref{table:features}: i) content-based textual features, ii) Wikipedia entity linking features and iii) video-based features. Although our dataset is composed of video lectures data, the majority of our features (with exception of some of the features in the video-based category) can be used across different modalities of educational material (e.g. books) as they are computed only considering the text transcription.
The transcriptions for the English lectures and the English translations of the non-English lectures are provided by the TransLectures project\footnote{\url{www.translectures.eu}}.

In this section, we define how different features are calculated from the lecture transcription. These features have been identified from the related work and are categorised under different verticals of quality assurance in text articles \cite{Bendersky2011,Dalip_wiki_svr,Ntoulas2006,wiki_wang} and engagement with video lectures \cite{Guo_vid_prod}. The verticals are for example understandability, topic coverage, presentation, freshness and authority \cite{quality_features}. The code for computing some of these features is available together with the dataset. 

\subsubsection{Content-based Features}
For explaining the features based on content transcripts, several functions need to be introduced: i) $\texttt{count}(s)$ is a function that returns the number of tokens in string $s$, ii) $\texttt{count}(t,s)$ is a function that returns the number of occurrences of tokens in token set $t$ in string $s$ and iii)
 $\texttt{u\_count}(t,s)$ returns the frequency of unique tokens from token set $t$ in string $s$. String $s$ can be the transcript text $s_{tr}$  or the lecture title $s_{title}$. \emph{Stop-word Presence Rate} and \emph{Stop-word Coverage Rate} are calculated using Eq. \ref{eq:swp} and \ref{eq:swc} based on the work of \citeauthor{Ntoulas2006}. Textual features defined by Eq. \ref{eq:prep} through Eq. \ref{eq:pron} are based on the work of \citeauthor{dalip_quality_features}. All definitions used the token sets provided in Table \ref{table:word_list}. 
More specifically, the content-based features extracted are the following: 
\begin{itemize}
    \item \emph{Word Count} of lecture transcript $s_{tr}$:
    \begin{equation}\label{eq:wc}
        \texttt{Word Count} = \texttt{count}(s_{tr})
    \end{equation}
    
    \item \emph{Title Word Count} of lecture $s_{title}$:
    \begin{equation}\label{eq:twc}
        \texttt{Title Word Count} = \texttt{count}(s_{title})
    \end{equation}
    
    \item \emph{Document Entropy}, based on the work of \citeauthor{Bendersky2011}, is calculated over every word $w$ in transcript $s_{tr}$ as: 
    \begin{equation}\label{eq:entropy}
        \texttt{Document Entropy} = \sum_{w \in s_{tr}} p_{s_{tr}}(w) \log p_{s_{tr}}(w),
    \end{equation}
    where $p_{s_{tr}}(w_i) = \frac{\texttt{count}(w_i,s_{tr})}{\texttt{Word Count}}$.
    
    \item \emph{FK Easiness} is computed using \texttt{textatistic} \cite{textatistic} for transcript $s_{tr}$ using:
    $\\\\ \texttt{FK Easiness} = $
    \begin{align}\label{eq:easiness}
         206.835 -
        1.015 \left(\frac{\texttt{Word Count}}{\texttt{sen\_count}(s_{tr})}\right) 
        - 84.6 \left(\frac{\texttt{syll\_count}(s_{tr})}{\texttt{Word Count}}\right)
    \end{align}
    where $\texttt{sen\_count}(s_{tr})$ and $\texttt{syll\_count}(s_{tr})$ returns the number of sentences and syllables in transcript $s_{tr}$ respectively. FK Easiness proxies complexity of the language used giving a low score for complex language and vice versa.
    \item \emph{Stop-word Presence Rate} of lecture transcript $s_{tr}$:
    \begin{equation}\label{eq:swp}
        \texttt{Stop-word Presence Rate} = \frac{\texttt{count}(sw, s_{tr})}{\texttt{Word Count}}
    \end{equation}
    
    \item \emph{Stop-word Coverage Rate} of lecture transcript $s_{tr}$:
    \begin{equation}\label{eq:swc}
        \texttt{Stop-word Coverage Rate} = \frac{\texttt{u\_count}(sw, s_{tr})}{\texttt{count}(sw)}
    \end{equation}
    
    \item \emph{Preposition Rate} of the lecture transcript $s_{tr}$:
    \begin{equation}\label{eq:prep}
        \texttt{Preposition Rate} = \frac{\texttt{count}(prep, s_{tr})}{\texttt{Word Count}}
    \end{equation}
    
    \item \emph{Auxiliary Rate} of the lecture transcript $s_{tr}$:
    \begin{equation}\label{eq:auxi}
        \texttt{Preposition Rate} = \frac{\texttt{count}(auxi, s_{tr})}{\texttt{Word Count}}
    \end{equation}
    
    \item \emph{To Be Rate} of lecture transcript $s_{tr}$:
    \begin{equation}\label{eq:tobe}
        \texttt{To Be Rate} = \frac{\texttt{count}(tobe, s_{tr})}{\texttt{Word Count}}
    \end{equation}
    
    \item \emph{Conjunction Rate} of lecture transcript $s_{tr}$:
    \begin{equation}\label{eq:conj}
        \texttt{Conjunction Rate} = \frac{\texttt{count}(conj, s_{tr})}{\texttt{Word Count}}
    \end{equation}
    
    \item \emph{Normalisation Rate} of lecture transcript $s_{tr}$:
    \begin{equation}\label{eq:norm}
        \texttt{Normalisation Rate} = \frac{\texttt{count}(norm, s_{tr})}{\texttt{Word Count}}
    \end{equation}
    
    \item \emph{Pronoun Rate} of lecture transcript $s_{tr}$:
    \begin{equation}\label{eq:pron}
        \texttt{Pronoun Rate} = \frac{\texttt{count}(pron, s_{tr})}{\texttt{Word Count}}
    \end{equation}
    
    \item \emph{Published Date} of video lecture $\ell$ calculates the epoch time of publication date of the lecture in days \cite{context_agnostic_engagement}:
    \begin{equation}\label{eq:age}
        \texttt{Published Date} = \texttt{days}(\ell_{pub\_date} - 1970/01/01)
    \end{equation}
\end{itemize}

Various prior works provide the rationale behind the suitability of these features \cite{quality_features,Guo_vid_prod,dalip_quality_features}.



\subsubsection{Wikipedia-based Features}

The Wikipedia topics most connected to the lectures are identified using Wikification \cite{wikifier}, an entity linking approach. Using the identified Wiki topics, four different feature groups are introduced with the dataset. They fall under the \emph{Authority} and \emph{Topic Coverage} verticals.

\begin{figure*}[h] 
  \centering
  \includegraphics[width=.9\linewidth]{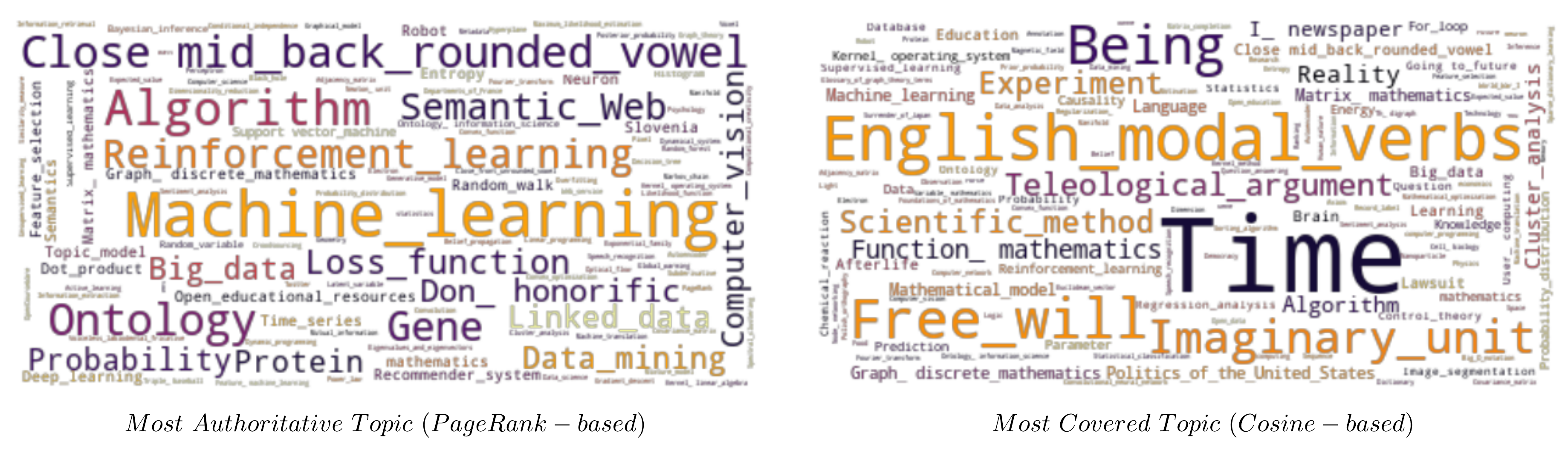}
  \caption{WordClouds summarising the 
  distribution of 
  the most authoritative (left) and most covered (right) Wikipedia topics in the dataset. Note that Computer Science and Data Science are the two dominant knowledge areas in our dataset.}
  \label{fig:wordcloud}
\end{figure*}

The \emph{top-5 authoritative topic URLs} and \emph{top-5 PageRank scores} features represent the Topic Authority feature vertical. Figure  \ref{fig:wordcloud} (left) shows the summary of Wikipedia topics that are most authoritative (top 1 topic) in the lectures found in the dataset. When PageRank score \cite{pagerank} is computed, Wikipedia topics heavily connected to other topics (i.e. more semantically related) within the lecture will emerge. Hence, the top-ranking topics are the more authoritative topics within the context of topics in the lecture.  
During Wikification \cite{wikifier}, a semantic graph is constructed where semantic relatedness ($SR(c,c')$) between each Wikipedia topic pair $c$ and $c'$ in the graph are calculated using:
\begin{equation}\label{eq:wiki_sr}
        SR(c,c') = \frac{\log(max(|L_c|, |L_{c'}|) - \log(|L_c \cap L_{c'}|)}
        {\log |W| - log (min(|L_c|, |L_{c'}|)}
\end{equation}
where $L_{c}$ represents the set of topics with inwards links to Wikipedia topic $c$, $|\cdot|$ represents the cardinality of the set and $W$ represents the set of all Wikipedia topics. This  semantic relatedness graph is used for computing PageRank scores. It is noteworthy that "authority" of a learning resource entails author, organisation and content authority \cite{quality_features}. These features represent content authority. The top 5 topic URLs and their relative PageRank Score are included as two feature groups providing 10 distinct features for each video lecture.

The \emph{top-5 covered topic URLs} and \emph{top-5 cosine similarity scores} features represent \emph{Topic Coverage} feature vertical. The cosine similarity score $cos(s_{tr}, c)$ between the \emph{Term Frequency-Inverse Document Frequency (TF-IDF)} representations of the lecture transcript $s_{tr}$ and the Wikipedia page $c$ is calculated using:
\begin{equation}\label{eq:wiki_cos}
        cos(s_{tr}, c) = \frac{\texttt{TFIDF}(s_{tr}) \cdot \texttt{TFIDF}(c)}
        {\|\texttt{TFIDF}(s_{tr})\| \times \|\texttt{TFIDF}(c)\|}
\end{equation}
where $\texttt{TFIDF}(s)$ returns the TF-IDF vector of string $s$.
Topics in the lecture are then ranked using this score.
Figure \ref{fig:wordcloud} (right) shows the summary of Wikipedia Topics that are most covered (top 1 topic) in the lectures found in the dataset.  The top 5 covered topic URLs and their cosine similarity scores are included as two additional feature groups providing 10 distinct features.

Topic authority and topic coverage features represent two different aspects of the content of a video lecture. Authoritative topics are the ones highly connected and dominant within the range of topics that are discussed in the lecture. An authoritative topic needs to have high semantic relatedness to other topics in the lecture. On the contrary, covered topics represent the heavy overlap between individual Wikipedia topics and the lecture transcript. Figure \ref{fig:wordcloud} gives further evidence of how these two feature groups are different from each other. The most emerging Wikipedia topics that are authoritative (left) in the lecture dataset are very different from the covered topics (right). The figure also shows that the authoritative topics are narrowly focused concepts (e.g. Machine Learning, Algorithm, Ontology, etc.) whereas the most covered topics tend to be more general topics (e.g. Time, Scientific Method, Unit, etc.).

\subsubsection{Video-specific Features} 
A set of easily automatable features that are video specific are  also included in the VLEngagement dataset. Features \emph{Lecture Duration, In Chunked, Lecture Type} and \emph{Speaker Speed} are calculated based on prior work \cite{Guo_vid_prod}. \emph{Lecture Duration} feature reports the duration of the video in seconds. \emph{Is Chunked} is a binary feature which reports\textit{True} if the lecture consists of multiple videos, and \textit{False} otherwise. \emph{Lecture type} value is derived from the metadata. The possible values for this feature are described in Table \ref{table:lect_type}.

A novel feature \emph{Silence Period Rate (SPR)} is introduced using the "silence" tags that are present in the video lecture transcript. The feature is defined as:
\begin{equation}\label{eq:SPR}
    \texttt{SPR}(\ell) =
    \frac{1}{D(\ell)}
    \sum_{t \in T(\ell)}
    D(t) \cdot \mathcal{I}(N(t) = {"silence"})
\end{equation}

where $t$ is a tag in the collection of tags $T(\ell)$ that belong to lecture $\ell$, $N$ returns the type of tag $t$ and $D$ returns the duration of tag $t$ or lecture $\ell$ and $\mathcal{I}(\cdot)$ is the indicator function (returning 1 when the condition is verified, 0 otherwise). 

\begin{table}
\small
\caption{14 types of lectures in the VLEngagement dataset and their abbriviation (Abbr.) and  frequency (Freq).}
\label{table:lect_type}
\centering
\begin{tabular}{ l l r l l r} \hline
    Abbr.&Description&Freq.&Abbr.&Description&Freq.\\ \hline
    \texttt{vbp}&Best Paper&16&\texttt{vdb}&Debate&30 \\
    \texttt{vdm}&Demonstration&124&\texttt{viv}&Interview&52 \\
    \texttt{vid}&Introduction&15&\texttt{vit}&Invited Talk&300 \\
    \texttt{vkn}&Keynote&115&\texttt{vl}&Lecture&2956 \\
    \texttt{vop}&Opening&31&\texttt{oth}&Other&15 \\
    \texttt{vpa}&Panel&44&\texttt{vps}&Poster&56 \\ 
    \texttt{vpr}&Promotional Video&23&\texttt{vtt}&Tutorial&269 \\ \hline
\end{tabular}
\end{table}

\begin{table}
\small
\caption{Features extracted and available in the VLEngagement dataset with their variable type (Continuous vs. Categorical) and their quality vertical.}
\label{table:features}
\centering
\begin{tabular}{l l c } \hline
    Type&Feature&Quality Vertical\\ \hline
   \multicolumn{3}{c}{\emph{Metadata features}} \\ \hline
    cat.&Language (English, non-English)&--- \\
    cat.&Domain (STEM, Miscellaneous)&--- \\ \hline
    \multicolumn{3}{c}{\emph{Content-based features}} \\ \hline
    
    con.&Word Count&Topic Coverage \\ 
    con.&Title Word Count&Topic Coverage\\ 
    con.&Document Entropy&Topic Coverage \\ 
    
    con.&Easiness (FK Easiness)&Understandability \\  
    con.&Stop-word Presence Rate&Understandability \\
    con.&Stop-word Coverage Rate&Understandability \\ 
    
    con.&Preposition Rate&Presentation \\ 
    con.&Auxiliary Rate&Presentation \\ 
    con.&To Be Rate&Presentation \\ 
    con.&Conjunction Rate&Presentation \\ 
    con.&Normalisation Rate&Presentation \\ 
    con.&Pronoun Rate&Presentation \\ 
    con.&Published Date&Freshness  \\ \hline
    
    \multicolumn{3}{c}{\emph{Wikipedia-based features}} \\ \hline 
    cat.&Top-5 Authoritative Topic URLs&Authority \\ 
    con.&Top-5 PageRank Scores &Authority \\ 
    cat.&Top-5 Covered Topic URLs&Topic Coverage \\ 
    con.&Top-5 Cosine Similarities&Topic Coverage \\  \hline

    \multicolumn{3}{c}{\emph{Video-based features}} \\ \hline    
    
    con.&Lecture Duration&Topic Coverage \\ 
    cat.&Is Chunked&Presentation \\ 
    cat.&Lecture Type&Presentation \\ 
    con.&Speaker speed&Presentation \\ 
    con.&Silence Period Rate (SPR)&Presentation \\ \hline

\end{tabular}
\end{table}

\subsection{Labels}

There are several target labels available in the VLEngagement dataset. These target labels are created by aggregating available explicit and implicit feedback measures in the repository. Mainly, the labels can be constructed as three different types of quantification's of learner subjective assessment of a video lecture. The relationship between these different subjective assessments metrics can be investigated with the VLengagement dataset.

\subsubsection{Explicit Rating} In terms of rating labels, \emph{Mean Star Rating} is provided for the video lecture using a star rating scale from 1 to 5 stars. As expected, explicit ratings are scarce and thus only populated in a subset of resources (1250 lectures). Lecture records are labelled with \texttt{-1} where star rating labels are missing. The data source does not provide access to ratings from individual users. Instead, only the aggregated average rating is available.

\subsubsection{Popularity} A popularity-based target label is created by extracting the \emph{View Count} of the lectures. The total number of views for each video lecture as of February 17, 2018 is extracted from the metadata and provided with the dataset.

\subsubsection{Watch Time/Engagement} The majority of learner engagement labels in the VLEngagement dataset are based on watch time. We aggregate the user view logs and use the Normalised Engagement Time (NET) to compute the \textbf{Median of Normalised Engagement (MNET)}, as it has been proposed as the gold standard for engagement with educational materials in previous work \cite{Guo_vid_prod}. We also calculate the \textbf{Average of Normalised Engagement (ANET)}. To have the MNET and ANET labels in the range $[0,1]$, we set the upper bound to 1 and derive Saturated MNET (\texttt{SMNET}) and Saturated ANET (\texttt{SANET}) respectively. Final \texttt{SMNET} (\emph{Median Engagement}) for lecture $\ell$ is computed as:
\begin{equation}
\label{eq:mnet}
    \texttt{SMNET}(\ell) = \max(\texttt{MNET}(\ell), 1)
\end{equation}

Similarly, \emph{Average Engagement} is calculated using:
\begin{equation}
\label{eq:anet}
    \texttt{SANET}(\ell) = \max(\texttt{ANET}(\ell), 1).
\end{equation}

The standard deviation of \texttt{NET} for each lecture (\emph{Std of Engagement}) is reported, together with the \emph{Number of User Sessions} used for calculating \texttt{MNET}. These additional features allow future studies to incorporate the degree of uncertainly and statistical confidence in the engagement labels (e.g. in their loss functions or performance metrics). Furthermore, the individual NET values for each lecture are also provided with the dataset. This allows having much more insight into the true distribution of NET for individual lectures rather than summary statistics. This data will allow future studies to refine engagement labels or use more sophisticated methods to predict engagement. 




\begin{table}
\small
\caption{Labels included in the VLEngagement dataset with their variable type, value interval and category.}
\label{table:labels}
\centering
\begin{tabular}{ l l l c } \hline
    Type&Label&Interval&Category\\ \hline
    cont.&Mean Star Rating & $[1,5)$ & Explicit Rating\\ 
    cont.&View Count & $(5,\infty)$ & Popularity\\ 
    
    cont.&SMNET (Eq. \ref{eq:mnet}) & $(0,1)$ & Watch Time \\ 
    cont.&SANET (Eq. \ref{eq:anet}) & $[0,1)$ & Watch Time \\ 
    
    cont.&Std. of NET & $(0,1)$ & Watch Time \\ 
    cont.&Number of User Sessions & $(5,\infty)$ & Watch Time \\ 
    cont.&Engagement Times (NET) & $[0,1)$ & Watch Time \\ \hline
\end{tabular}
\end{table}

\subsection{Anonymity}
We restrict the final dataset to lectures that have been viewed by at least 5 unique users to have reliable engagement measurements. Additionally, a regime of techniques are used for preserving the anonymity of the lectures  in order to preserve the identities of the authors/lecturers. 
The motivation behind this decision is to avoid authors of the video lectures having unanticipated effects on their reputation by associating implicit learner engagement values to their content.

Rarely occurring values in \emph{Lecture Type} feature were grouped together to create the \emph{other} category found in Table \ref{table:lect_type}. \emph{Language} feature is grouped into \texttt{en} and \texttt{non-en} categories. Similarly, Domain category groups Life Sciences, Physics, Technology, Mathematics, Computer Science, Data Science and Computers subjects to \texttt{stem} category and the other subjects to \texttt{misc} category. Rounding is used with \emph{Published Date}, rounding to the nearest 10 days. \emph{Lecture Duration} is rounded to the nearest 10 seconds. Gaussian white noise (10\%) is added to \emph{Title Word Count} feature and rounded to the nearest integer.
 
\subsection{Final Dataset}
 The final dataset includes lectures that are 
 published between September 1, 1999 and October 1, 2017. The 
 engagement labels are created from 155,850 user views logged between December 8, 2016 and February 17, 2018. 
 The final dataset consists of 4,046 lectures across 21 subjects (eg.~Computer Science, Philosophy, etc.) that are categorised into STEM and Miscellaneous domains. The dataset, helper tools and example code snippets are available publicly\footnote{\url{https://github.com/sahanbull/context-agnostic-engagement}}
 .


\section{Supported Tasks} \label{sec:taks}

This section introduces the reader to the tasks that the dataset could be used for. The main application areas of these tasks are quality assurance in open education and scientific content recommenders and understanding and predicting population engagement in an online learning setting. Tasks 1 and 2 are demonstrated in this paper. Tasks 3-6 have been partially tackled in our prior work \cite{context_agnostic_engagement}. Tasks 7-8 are novel.

We establish two main tasks, which we mainly focus on in this paper, that can be objectively addressed using the VLEngagement dataset using a supervised learning approach. These are:
\begin{enumerate}
    \item \textbf{Task 1: Predicting context-agnostic (population-based) engagement of video lectures}: The dataset provides a set of relevant features and labels to construct machine learning models to predict context-agnostic engagement in video lectures. 
    The task can be treated as a regression problem to predict the different engagement labels.
    \item \textbf{Task 2: Ranking of video lectures based on engagement}: Building predictive models that could rank lectures based on their context-agnostic engagement could be useful in the setting of an educational recommendation system, including tackling the cold-start problem associated to new video lectures. The task can be treated as a ranking problem to predict the global/relative ranking of video lectures. 
\end{enumerate}

We further identify several auxiliary tasks that can also be addressed with this dataset:
\begin{itemize}
    \item \textbf{Task 3: Features influencing engagement}: Uncovering the role of different textual and video-specific features involved in several statistics of population-based engagement. 
    \item \textbf{Task 4: Influence of topics in engagement}: Understand the role that the topical content in the lecture play on population based engagement (with link to the Wikipedia pages of these topics). 
    \item \textbf{Task 5: Disentangle different factors from engagement}: Compare features involved in engagement for different video lecture types, language and knowledge areas (e.g. STEM vs non-STEM lectures). 
    \item \textbf{Task 6: Comparing different measures of implicit and explicit subjective assessment}: Analyse the differences between engagement vs mean star ratings and number of views to identify the strengths and weaknesses of the different feedback types. 
    \item \textbf{Task 7: Unsupervised learning to understand the distribution of video lectures}: Cluster video lectures according to the provided features to understand their distribution. Identification of formal patterns that depict similarities and differences between lectures could be insightful.  
   \item \textbf{Task 8: Deducing the structure of knowledge}: The co-occurrence patterns of topics within the video lectures provide a great source of data to understand inter-topic relationships and how knowledge is structured. Work in this direction can be used in identifying related materials and accounting for novelty in educational recommendation \cite{truelearn}. 
   \item \textbf{Task 9: Contrasting to other educational datasets}: The lectures in the VLEngagement dataset are scientific videos, thus it may be meaningful to study if similar patterns for engagement hold across other educational datasets that come from other settings (e.g.: MOOCs). 
\end{itemize}

We propose two baseline models addressing the main tasks (1 and 2) in section \ref{sec:baselines}. 




\subsection{Evaluating Performance}
 We identify \emph{Root Mean Squared Error (RMSE)} as a suitable metric for Task 1. Measuring RMSE against the original labels published with the datasets will allow different works to be compared fairly. With reference to Task 2, we identify \emph{Spearman's Rank Order Correlation Coefficient (SROCC)} and \emph{Pairwise Ranking Accuracy (Pairwise)}. SROCC 
 is suitable for comparing between ranking models that create global rankings (e.g. point-wise ranking algorithms). However, pairwise ranking accuracy is more intuitive for this task as it represents the fraction of pairwise comparisons where the model could predict the more engaging lecture. There is more than one unique solution for this problem, especially when there is error associated with the ranking model \cite{Furnkranz2011}.

 We use 5-fold cross validation to evaluate model performance with tasks 1 and 2. We release the folds together with the dataset, to allow for fair comparisons to the baselines. The five folds can be identified using the \texttt{fold} column in the dataset. 
 5-fold cross validation also allows reporting the \emph{standard error (1.96 $\times$ Standard Deviation)} of the performance estimate, which we include in our results in tables \ref{table:rmse} and \ref{table:accuracy}.
 
\section{Experiments and Baselines} \label{sec:baselines}

Prior work on similar tasks identify ensemble models \cite{context_agnostic_engagement,wiki_wang} to be the best performing models with the main tasks described in section \ref{sec:taks}.
We use \emph{Random Forests Regressor (RF)} and \emph{Gradient Boosting Machines (GBM)for constructing baselines.} We use \emph{SMNET} labels as the target variable for both engagement prediction and video lecture ranking tasks. No pre-processing or cleaning steps are necessary.

\subsection{Features and Labels for Baseline Models}

All the features outlined in the content-based and video-based sections in Table \ref{table:features} are included in the baseline models. However, due to the large amount of topics available in the Wikipedia-based feature groups, we restrict the feature set by adding only the \emph{most authoritative topic URL} and \emph{most covered topic URL}, where both the features are added to the baseline models as categorical variables. Practitioners are encouraged to try further encodings of these variables, as it will likely have a great impact in the performance.

The models are trained with three different feature sets in an incremental fashion: 
\begin{enumerate}
    \item \emph{Content-based}: Features extracted from lecture metadata and the textual features extracted from the lecture transcript. 
    \item \emph{+ Wiki-based}: In addition to the content-based features, two Wikipedia based features (most authoritative topic URL and most covered topic URL) are added to the feature set.
    \item \emph{+ Video-based}: In addition to both content-based and Wikipedia-based features, video specific features are added.
\end{enumerate}
This allows identifying the performance gain achieved through adding each new group of features.

Our preliminary investigations indicated that SMNET label follows a Log-Normal distribution, motivating us to use a log transformation on the SMNET values before training the models. Empirical results further confirmed that this step improves the final performance of the models. We undo this transformation for computing $RMSE$.    

\subsection{Results and Discussion}

The results for engagement prediction task (Task 1) are reported in Table \ref{table:rmse}. Table \ref{table:accuracy} reports the performance in ranking lectures based on engagement (Task 2). It is evident that addition of Wikipedia-based features and video-specific features contribute towards improving model performance across both tasks with video-specific features leading to significant gains. The results show that the RF model is consistently better at predicting lecture engagement (Table \ref{table:rmse}) whereas the GBM model dominates the performance in lecture ranking (Table \ref{table:accuracy}) although these two models belong to the ensemble learning family. 

This dataset provides us with the opportunity to understand context-agnostic engagement with a unique type of video lectures, specifically, scientific videos. Although the results in tables \ref{table:rmse} and \ref{table:accuracy} show that adding Video-specific features leads to consistent improvements of predictive performance, it is evident that the cross-modal content-based features alone lead to substantial amount of predictive performance in comparison to the gains by adding modality-specific features. This is a good indication that easy-to-compute, cross-modal features alone are sufficient to build a system that can predict context-agnostic engagement of video lectures to a satisfactory degree.

The results also indicate that there is no significant gain in performance by adding the Wikipedia features. However, we believe that this is due to the simplicity of the Wiki features used in constructing the baselines.


\subsection{Limitations and Opportunities} \label{limitations}
This dataset has several limitations that are noteworthy. For example, as the topics in Figure \ref{fig:wordcloud} indicate, this dataset is dominated with Computer Science and Data Science related lectures that are mainly delivered in English. In addition, the majority of lectures in the dataset are research talks, narrowing down the style and type of data. 
These limitations cast significant uncertainty regarding the generalisation of the prediction models to more diverse types of educational video lectures. Although VLEngagement dataset is large compared to the rest of educational engagement datasets available, it still suffers from a limitation in the variety of its data.

\emph{Learner Engagement} is a loaded concept with many facets. In relation to consuming videos, many behavioural actions such as pausing, rewinding and skipping can contribute to latent engagement with a video lecture \cite{lan2017behavior}. Analysing facial expressions and affective states is another alternative approach to representing engagement \cite{dewan2019engagement}. However, due to the technical limitations of the platform and privacy concerns, only watch time, number of  views and mean ratings are included in this dataset. Although watch time has been used as a representative proxy for learner engagement with videos \cite{Guo_vid_prod,beyondviews}, we acknowledge that more informative measures may lead to more complete and reliable engagement signals.

Although this is the case, there are numerous opportunities that are presented by this dataset. It provides the opportunity to understand engagement with scientific videos and to what extent the engagement dynamics align/differ with other types of educational videos. In addition to the summarised engagement signals, the individual user engagement signals are provided with the dataset. This data will allow researchers to better understand the engagement distribution and apply more creative techniques to flesh out the engagement signals.

\begin{table}[]
 \small
\caption{Test RMSE for the engagement prediction models (task 1) with standard error (lower values are better).}\label{table:rmse}
\begin{tabular}{ l l l }
\hline
& \multicolumn{2}{c}{RMSE} \\
Feature Set & \multicolumn{1}{c}{GBM} & \multicolumn{1}{c}{RF} \\ \hline
Content-based &.1802$\pm$.0160 & \textbf{.1801$\pm$.0137} \\ 
+ Wiki-based & .1814$\pm$.0160 & \textbf{.1798$\pm$.0148}  \\ 
+ Video-specific & .1737$\pm$.0172 & \textbf{.1728$\pm$.0160} \\ \hline
\end{tabular}
\end{table}

\begin{table}[t]
\small
 \setlength{\tabcolsep}{2pt}
\caption{Test SROCC and Pairwise Ranking Accuracy (Pairwise) for lecture ranking models (task 2) with standard error (higher values are better).}\label{table:accuracy}
\begin{tabular}{ l l l l l }
\hline
\multicolumn{1}{r}{Model} & \multicolumn{2}{c}{GBM} & \multicolumn{2}{c}{RF} \\

Feature Set & \multicolumn{1}{c}{SROCC} & \multicolumn{1}{c }{Pairwise} & \multicolumn{1}{c }{SROCC} & \multicolumn{1}{c}{Pairwise} \\ \hline

Content-based & \textbf{.6241$\pm$.0291} & \textbf{.7221$\pm$.0102} & .6190$\pm$.0237 & .7202$\pm$.0086  \\ 
+ Wiki-based & .6245$\pm$.0339 & .7224$\pm$.0115 & \textbf{.6251$\pm$.0322} & \textbf{.7225$\pm$.0123}  \\ 
+ Video-specific & \textbf{.6761$\pm$.0434} & \textbf{.7446$\pm$.0183} & .6758$\pm$.0458 & .7446$\pm$.0197  \\ \hline
\end{tabular}
\end{table}

\begin{table*}
\small
\caption{Tokens used for Feature Extraction.}
\label{table:word_list}
\centering
\begin{tabular}{ l c p{13cm} } \hline
    Token Set&Description&Tokens\\ \hline
    \texttt{sw}&Stopwords&all, show, anyway, fifty, four, go, mill, find, seemed, one, whose, re, herself,  
 whoever, behind, should, to, only, under, herein, do, his, get, very, de, none, 
    cannot, every, during, him, did, cry, beforehand, these, she, thereupon, where,
    ten, eleven, namely, besides, are, further, sincere, even, what, please, yet, couldn\'t,
    enough, above, between, neither, ever, across, thin, we, full, never, however, here,
     others, hers, along, fifteen, both, last, many, whereafter, wherever, against, etc, s,
     became, whole, otherwise, among, via, co, afterwards, seems, whatever, alone,
     moreover, throughout, from, would, two, been, next, few, much, call, therefore,
     interest, themselves, thr, until, empty, more, fire, latterly, hereby, else,
    everywhere, former, those, must, me, myself, this, bill, will, while, anywhere,
     nine, can, of, my, whenever, give, almost, is, thus, it, cant, itself, something, in, 
     ie, if, inc, perhaps, six, amount, same, wherein, beside, how, several, whereas, see,
      may, after, upon, hereupon, such, a, off, whereby, third, i, well, rather, without, so,
     the, con, yours, just, less, being, indeed, over, move, front, already, through,
     yourselves, still, its, before, thence, somewhere, had, except, ours, has, might,
     thereafter, then, them, someone, around, thereby, five, they, not, now, nor, 
     name, always, whither, t, each, become, side, therein, twelve, because, often, 
     doing, eg, some, back, our, beyond, ourselves, out, for, bottom, since, forty, per,
     everything, does, three, either, be, amongst, whereupon, nowhere, although,
     found, sixty, anyhow, by, on, about, anything, theirs, could, put, keep, whence,
     due, ltd, hence, onto, or, first, own, seeming, formerly, into, within, yourself, 
     down, everyone, done, another, thick, your, her, whom, twenty, top, there, system, 
     least, anyone, their, too, hundred, was, himself, elsewhere, mostly, that, becoming,
     nobody, but, somehow, part, with, than, he, made, whether, up, us, nevertheless,
     below, un, were, toward, and, describe, am, mine, an, meanwhile, as, sometime, 
     at, have, seem, any, fill, again, hasn\'t, no, latter, when, detail, also, other, take, 
    which, becomes, yo, towards, though, who, most, eight, amongst, nothing,
     why, don, noone, sometimes, together, serious, having, once, hereafter \\ \hline 
    \texttt{conj}&Conjunctions&and, but, or, yet, nor
    \\ \hline 
    \texttt{norm}&Normalizations&-tion, -ment, -ence, -ance
     \\ \hline 
    \texttt{tobe}&To-be Verbs&be, being, was, were, been, are, is
     \\ \hline 
    \texttt{prep}&Prepositions&aboard, about, above, according to, across from, after, against, alongside,
     alongside of, along with, amid, among, apart from, around, aside from, at,
     away from, back of, because of, before, behind, below, beneath, beside,
     besides, between, beyond, but, by means of, concerning, considering, despite, 
     down, down from, during, except, except for, excepting for, from among, 
     from between, from under, in addition to, in behalf of, in front of, in place of, 
     in regard to, inside of, inside, in spite of, instead of, into, like, near to, off,
     on account of, on behalf of, onto, on top of, on, opposite, out of, out, outside, 
     outside of, over to, over, owing to, past, prior to, regarding, round about,
     round, since, subsequent to, together, with, throughout, through, till, toward,
     under, underneath, until, unto, up, up to, upon, with, within, without, across, 
     long, by, of, in, to, near, of, from
    \\ \hline 
    \texttt{auxi}&Auxiliary Verbs&will, shall, cannot, may, need to, would, should, could, might, must, ought, ought to, can’t, can
    \\ \hline 
    \texttt{pron}&Pronouns&i, me, we, us, you, he, him, she, her, it, they, them, thou, thee, ye, myself,
    yourself, himself, herself, itself, ourselves, yourselves, themselves, oneself, 
    my, mine, his, hers, yours, ours, theirs, its, our, that, their, these, this, those
    \\ \hline 
\end{tabular}
\end{table*}

\section{Conclusions and Future Directions}

Identifying the need for understanding context-agnostic engagement prediction to improve scalable quality assurance and recommendations systems in education, we have constructed and published a novel dataset with a wide range of features for over 4000 scientific video lectures. The dataset consists of a diverse set of lectures belonging to multiple languages, knowledge areas and lecture types with features that are content-based, Wikipedia-based and video specific. In the spirit of improving engagement prediction in video lectures, we establish two main tasks, (i) predicting context-agnostic engagement of video lectures and (ii) ranking video lectures based on engagement, together with 7 auxiliary tasks that can be addressed with this dataset. Ensemble learning methods tend to perform well in this task, leading to introducing two baseline models for the two main tasks. The promising performance of the models with the dataset demonstrates the possibility of building machine learning models to predict engagement in video lectures.    

We plan several lines of future work relating to improving the limitations of the current version of the dataset (and therefore the potential tasks it can be used for). This entails both horizontal and vertical expansion of the dataset. Horizontal expansions relates to introducing new features. More content-based features can be computed by exploiting the semantic graph constructed with the Wikipedia topics \cite{ponza2020computing}.
A wider range of features that capture textual, audio-visual and presentation slides related patterns will be constructed \cite{edx_quality}.
Computer vision based features for videos and processing visual information in educational material (slides in videos) can be provided to improve modality-specific feature sets. Vertical expansions of the dataset relate to adding new observations. Adding more video lectures coming from multiple sources such as YouTube would widen the diversity of data.
Following the reflections from section \ref{limitations}, the possibility of including more learner engagement related signals (e.g.: pauses, replays, skips, etc.) will be explored in the subsequent version of the dataset, without compromising learner privacy.
As more understanding of engagement with other modalities (such as PDFs and e-Books) is gained, it is possible to add more observations from diverse modalities to widen the horizons of the dataset and improve understanding of engagement with different modalities of educational material. Additional features with more diverse observations and representations may unlock the possibility of experimenting with more sophisticated deep learning and multi-task learning models. We will also connect the dataset to learners' personalised data through our future work in order to support building personalised tasks and making the connection to population-based engagement, which has been suggested in previous work as an important step towards building integrative educational recommender systems \cite{trueeducation}.


\begin{acks}
This research is part of the EU's Horizon 2020 
grant No 761758 (\url{www.x5gon.org}) and partially funded by the EPSRC Fellowship titled "Task Based Information Retrieval", under grant No EP/P024289/1.
\end{acks}

\bibliographystyle{ACM-Reference-Format}
\bibliography{sample-base}










\end{document}